\documentclass[11pt]{article}
\usepackage{graphicx,amsmath,amsthm,mathrsfs,amssymb}
\usepackage[usenames]{color}
\usepackage{ulem}
\usepackage{authblk}
\usepackage{cite}
\newcommand{\comm}[1]{{\color[rgb]{1,0,0}\small  Comment: #1}}

\newcommand{\ee}{\end{equation}}
\newcommand{\word}[1]{\,\,\mbox{#1}\,\,}
\newcommand{\reff}[1]{(\ref{#1})}
\newcommand{\beq}{\begin{equation}}
\newcommand{\eeq}[1]{\label{#1}\end{equation}}
\newcommand{\beqa}{\begin{eqnarray}}
\newcommand{\eea}{\end{eqnarray}}
\newcommand{\eeqa}[1]{\label{#1}\end{eqnarray}}
\newcommand{\beg}{\begin{equation*}}
\newcommand{\eeg}{\end{equation*}}

\newcommand{\m}{\!-\!}

\newcommand{\bsplit}{\begin{split}}
\newcommand{\esplit}{\end{split}}

\title{Gravitating magnetic monopole \\ via the spontaneous symmetry breaking of pure $R^2$ gravity}

\author[1]{Ariel Edery\thanks{corresponding author: aedery@ubishops.ca}}
\author[2]{Yu Nakayama\thanks{yu.nakayama@rikkyo.ac.jp}}

\affil[1]{Department of Physics, Bishop's University, \\2600 College Street, Sherbrooke, Qu\'{e}bec, Canada, J1M 1Z7 .\vspace{8mm}}
\affil[2]{
Department of Physics, Rikkyo University, Toshima, Tokyo 171-8501, Japan}
\begin{document}
\date{}
\maketitle
\begin{abstract} 
The pure $R^2$ gravity is equivalent to Einstein gravity with cosmological constant and　
a massless scalar field, and it further possesses the so-called restricted Weyl symmetry, which is a symmetry larger than scale symmetry. To incorporate matter, we consider a restricted Weyl invariant action composed of pure $R^2$ gravity, SU(2) Yang-Mills fields, and a non-minimally coupled massless Higgs field (a triplet of scalars). When the restricted Weyl symmetry is spontaneously broken, it is equivalent to an Einstein-Yang-Mills-Higgs (EYMH) action with a cosmological constant and a massive Higgs non-minimally coupled to gravity i.e. via a term  $\tilde{\xi} R |\vec{\Phi}|^2$. When the restricted Weyl symmetry is not spontaneously broken, linearizations about Minkowski spacetime do not yield gravitons in the original $R^2$ gravity and hence it does does not gravitate. However, we show that in the broken gauge sector of our theory, where the Higgs field acquires a non-zero vacuum expectation value, Minkowski spacetime is a viable gravitating background solution. We then obtain numerically gravitating magnetic monopole solutions for non-zero coupling constant $\tilde{\xi}=1/6$ in three different backgrounds: Minkowski, anti-de Sitter (AdS) and de Sitter (dS), all of which are realized in our restricted Weyl invariant theory.
\end{abstract}

\setcounter{page}{1}
\section{Introduction}
In the last few years there has been an interest in studying pure $R^2$ gravity (i.e. gravity action solely made out of $R^2$ with no additional $R$ term or cosmological constant, where $R$ is the Ricci scalar). This interest arises from the fact that this theory is equivalent to Einstein gravity with non-zero cosmological constant and a massless scalar field as recently discussed in \cite{Lust1,Lust2,YNAE2,Lust3}. Despite being a higher-derivative gravity theory, this equivalence implies that the pure $R^2$ gravity is unitary in contrast to all the other quadratic gravity theories like the square of the Ricci tensor, Riemann tensor or Weyl tensor\cite{Lust3}. It has been known for a long time that this theory is invariant under scale transformations $g_{\mu\nu}\to \lambda^2 \,g_{\mu\nu}$ where $\lambda$ is a constant. However, it has recently been pointed out that it has a symmetry, called restricted Weyl symmetry, which is larger than scale symmetry \cite{YNAE1}. It is invariant under the Weyl transformation $g_{\mu\nu} \to \Omega^2(x) \, g_{\mu\nu}$ where the conformal factor $\Omega(x)$ obeys the condition $\Box\,\Omega(x)\equiv g^{\mu\nu}\nabla_{\mu}\nabla_{\nu}\Omega(x)=0$. Clearly, $\Omega(x)$  is not limited to being a constant. 

The equivalence of the $R^2$ gravity with Einstein gravity occurs when the restricted Weyl symmetry is spontaneously broken. This happens when the background (vacuum) spacetime has $R\neq 0$ which includes de Sitter (dS) and anti-de Sitter(AdS) spacetime but not Minkowski spacetime. The symmetric $R=0$ vacuum corresponding to the unbroken sector, is a completely separate case and has no correspondence to Einstein gravity. In fact, it has been shown that linearization about Minkowski spacetime yields a propagating scalar but no propagating graviton; about a Minkowski background, pure $R^2$ gravity does not gravitate at the linearized level \cite{Lust3}. 

In this paper, we consider a restricted Weyl invariant action that includes $R^2$ gravity,  $SU(2)$ Yang-Mills fields, and a massless Higgs field $\vec{\Phi}$ (a triplet of scalars) non-minimally coupled to gravity. Note that the Higgs field must be massless here to maintain the restricted Weyl symmetry. We then show that this action is equivalent to an action that includes Einstein gravity, a non-zero cosmological constant, a non-minimally coupled Higgs field  and a massless scalar field $\varphi$ (redefined later in terms of a canonically normalized field $\psi$). In going over to the Einstein action, the Higgs field acquires a mass via the spontaneous breaking of the restricted Weyl symmetry. The original Yang-Mills fields are also present. We therefore obtain an Einstein-Yang-Mills-Higgs (EYMH) action plus extra terms. The extra terms include a cosmological constant, a non-minimal coupling term $R\,|\vec{\Phi}|^2$, a massless scalar field $\varphi$ and an interaction term between $\varphi$ and the Higgs field. This action yields gravitating magnetic monopole (and black hole) solutions in de Sitter (dS), anti-de Sitter (AdS) and Minkowski backgrounds. 

In pure $R^2$ theory without the Higgs field, $R=0$ corresponds to the symmetric vacuum (the unbroken sector) where Minkowski space is not an interesting background: as previously mentioned, linearizations about it do not yield gravitons. However, when a non-minimal coupling to the Higgs field is included in the $R^2$ action, and the Higgs field acquires a non-zero vacuum expectation value (VEV), the restricted Weyl symmetry as well as the gauge symmetry are spontaneously broken. In this case, $R=0$ together with $\vec{\Phi} \neq 0$ is in the broken sector and is therefore a valid background in the equivalent Einstein action. Of course, it is well known that in Einstein gravity linearizations about Minkowski spacetime lead to gravitational waves i.e. in Einstein gravity Minkowski spacetime gravitates at the linearized level. In the original $R^2$ action, the non-minimal coupling to a massless Higgs field with non-zero VEV plays therefore two crucial roles: it yields magnetic monopole solutions when Yang-Mills fields are present but it also provides $R^2$ gravity with a viable Minkowski background. 

EYMH monopoles (and black holes) were originally studied in Minkowski backgrounds \cite{Maison,Ortiz,Weinberg} and later in curved backgrounds \cite{Lugo,Lugo2,Maison2,Li}. These studies were carried out with the Higgs field minimally coupled to the Ricci scalar.  As already stated, our final action is an EYMH action that contain extra terms such as a non-zero cosmological constant and a non-minimal coupling to the Higgs field. These extra terms are a natural outcome of the restricted Weyl invariance of the $R^2$ action. The monopoles studied in this paper have Minkowski as well as AdS and dS backgrounds but contain a non-minimal coupling term $\tilde{\xi} R |\vec{\Phi}|^2$ with coupling constant $\tilde{\xi}=1/6$ (the reason for this value will be explained later). We obtain numerical solutions for magnetic monopoles non-minimally coupled in all three backgrounds and plot the solutions in each case. We also obtain analytical Reissner-Nordstr\"om black hole solutions in Minkowski, dS and AdS backgrounds. By definition, the black hole has a singularity at the origin whereas the magnetic monopoles are regular at the origin. 

It is worth mentioning that the static spherically symmetric Einstein-Yang-Mills equations without Higgs field does contain a particle-like soliton (non-singular solution), often referred to as the Bartnik-McKinnon solution \cite{Bartnik}. However, in contrast to the EYMH monopole, it has no global charge. Cosmological analogs of these particle-like solutions in dS spacetime have also been found \cite{Volkov}. A good review of a large class of compact objects in Einstein-Yang-Mills theories can be found in \cite{Lemos}. Magnetic monopoles under gravity other than Einstein gravity have also been studied in \cite{Fabbri, Noah}.
      

\section{Restricted Weyl invariant $R^2$ action with Higgs field and its equivalent Einstein action}
We begin with an action that includes pure $R^2$ gravity, a non-minimally coupled massless triplet Higgs field $\Phi$ and $SU(2)$ non-abelian gauge fields $A^i_{\mu}$: 
\begin{align}
S_a = \int d^4x \sqrt{-g} \left( \alpha R^2 - \xi R |\vec{\Phi}|^2 - D_\mu \vec{\Phi} D^\mu \vec{\Phi} - \lambda |\vec{\Phi}|^4 + \frac{1}{4}\mathrm{Tr}F_{\mu\nu} F^{\mu\nu} \right) 
 \label{Rw1}
\end{align}
where $\alpha$, $\xi$ and $\lambda$ are free dimensionless parameters and $D_{\mu}$ is the usual covariant derivative with respect to the non-abelian gauge symmetry. The above action is restricted Weyl invariant \cite{YNAE1,YNAE2} i.e. it is invariant under the transformation 
\beq
g_{\mu\nu}\rightarrow \Omega^2 g_{\mu\nu}\quad,\quad \vec{\Phi}\rightarrow \vec{\Phi}/\Omega\quad,\quad A^i_{\mu} \rightarrow A^i_{\mu} \word{ with} \Box \Omega=0
\eeq{Rw2}
where the conformal factor $\Omega(x)$ is a real smooth function. This symmetry forbids a mass term for the Higgs, an Einstein-Hilbert term as well as a cosmological constant term in \reff{Rw1}. Introducing the auxiliary field $\varphi$, we can rewrite the above action into the equivalent form 
\begin{align}
S_b =\int d^4x \sqrt{-g} \Big[-\alpha(c_1 \varphi + R + &\frac{c_2}{\alpha} |\vec{\Phi}|^2)^2 + \alpha R^2 \m \xi R |\vec{\Phi}|^2 \cr &- D_\mu \vec{\Phi} D^\mu \vec{\Phi} -\lambda |\vec{\Phi}|^4 + \frac{1}{4}\mathrm{Tr}F_{\mu\nu} F^{\mu\nu} \Big]  
\label{Sb}
\end{align} 
where $c_1$ and $c_2$ are arbitrary constants\footnote{The squared term yields a Gaussian integral over $\varphi$ in the path integral and does not affect anything i.e. $\int\mathcal{D}\varphi e^{- i\alpha\,c_1^2 \int d^4x\sqrt{-g}(\varphi - f(x))^2} = \mathrm{const}$.}. Expanding the above action we obtain  
\begin{align}
S_c &= \int d^4x \sqrt{-g} \Big(-c_1^2\,\alpha \,\varphi^2 - 2\alpha c_1 \varphi R    -(\xi+2c_2)R|\vec{\Phi}|^2  \cr
& \qquad\qquad -  D_\mu \vec{\Phi} D^\mu \vec{\Phi}-2c_1c_2 \varphi |\vec{\Phi}|^2  -(\alpha^{-1} c_2^2 +\lambda)|\vec{\Phi}|^4  +\frac{1}{4}\mathrm{Tr}F_{\mu\nu} F^{\mu\nu} \Big) \ . 
\label{Sc}
\end{align}
Action \reff{Sc} is equivalent to the original action \reff{Rw1} and is restricted Weyl invariant as long as $\varphi$ transforms accordingly; it is invariant under the transformations $g_{\mu\nu}\rightarrow \Omega^2 g_{\mu\nu}$, $\varphi\rightarrow \varphi/\Omega^2$, $\vec{\Phi}\rightarrow \vec{\Phi}/\Omega$, $A^i_{\mu} \rightarrow A^i_{\mu}$ with $\Box \Omega=0$. 

After performing the conformal (Weyl) transformation
\beq
g_{\mu\nu} \to \varphi^{-1}g_{\mu\nu}\quad,\quad\vec{\Phi} \to \varphi^{1/2} \vec{\Phi} \quad,\quad A^i_{\mu} \to A^i_{\mu}
\eeq{Conf}
the above action reduces to an Einstein-Hilbert action with a massive Higgs term plus other terms   
\begin{align}
S_d=\int d^4x \sqrt{-g} \,\Big(&\m\alpha c_1^2 \m 2\,\alpha c_1 R \m  D_\mu \vec{\Phi} D^\mu\vec{\Phi} \m 2\,c_1c_2 |\vec{\Phi}|^2 \m (\alpha^{-1} c_2^2 +\lambda)|\vec{\Phi}|^4  -(\xi+2c_2)R\,|\vec{\Phi}|^2 \nonumber\\& + 3\,\alpha c_1 \dfrac{1}{\varphi^2} \partial_\mu \varphi \,\partial^\mu \varphi +(6(\xi+2c_2)-1)\varphi^{1/2} \Box \varphi^{-1/2}|\vec{\Phi}|^2 + \frac{1}{4}\mathrm{Tr}F_{\mu\nu} F^{\mu\nu}\Big).
\label{Sd}
\end{align} 
The constants $-2 \alpha c_1$ and $-\alpha \,c_1^2$ determine Newton's constant and the cosmological constant respectively and can be chosen freely by adjusting the parameters $\alpha$ and $c_1$ (with $\alpha \,c_1 <0$ to ensure the correct sign for Newton's constant)\footnote{The constant $c_1$ is dimensionful and has units of (length)$^{-2}$. This stems from the fact that $c_1 \varphi$ in \reff{Sb} has units of (length)$^{-2}$ and $\varphi$ is assumed dimensionless in \reff{Conf}. In contrast, the constant $c_2$ is dimensionless.}.  We are then free to choose $c_2$, $\lambda$ and $\xi$ to fix the Higgs mass, the coefficient $-(\alpha^{-1} c_2^2 +\lambda)$ of the quartic term and the coefficient of the non-minimal coupling term $R \,|\vec{\Phi}|^2$ respectively. By defining $\psi =\sqrt{-3\,\alpha c_1}\ln \varphi$ the kinetic term for $\varphi$ can be expressed in the canonical form $-\partial_\mu \psi \,\partial^\mu \psi$.  

The conformal (Weyl) transformation \reff{Conf} is not valid for $\varphi=0$ and therefore the possibility of the Weyl transformation and the equivalence of the theories tacitly assumes a vacuum with $\varphi \neq 0$. This vacuum is not invariant under $\varphi \to \varphi/\Omega^2$ so that the restricted Weyl symmetry is spontaneously broken. This is evident from the fact that the final action \reff{Sd} now has a massive Higgs and Einstein-Hilbert term. The massless scalar $\psi$ (defined above in terms of $\varphi$), is identified as the Nambu-Goldstone boson associated with the broken symmetry. It is well known that in spontaneously broken theories the original symmetry is still realized as a shift symmetry of the Goldstone bosons \cite{Schwartz}. This is the case here. The action \reff{Sd} is invariant under $\varphi\to \varphi/\Omega^2$, $g_{\mu\nu}\to g_{\mu\nu}$, $\vec{\Phi} \to \vec{\Phi}$ with condition\footnote{The original restricted Weyl symmetry (with metric tensor denoted with a hat) required the condition  $\hat{\Box} \Omega =0$. This condition, after the replacement $\hat{g}_{\mu\nu} = \varphi^{-1} g_{\mu\nu}$ becomes $\Box \Omega -\partial_\mu (\ln \varphi) \partial^\mu \Omega =0$. Note that $\varphi^{1/2} \Box \varphi^{-1/2}$ is invariant under $\varphi \to \varphi/\Omega^2$ when the condition $\Box \Omega -\partial_\mu (\ln \varphi) \partial^\mu \Omega =0$ holds.} $\Box \Omega -\partial_\mu (\ln \varphi) \partial^\mu \Omega =0$. The Goldstone boson $\psi$ therefore undergoes the shift symmetry $\psi \to \psi - 2 \sqrt{-3\,\alpha c_1}\ln \Omega$. Note that the shift symmetry not only forbids a mass term for $\psi$ but places constraints on how it couples to the Higgs field; it determines the nature of the interaction term between $\varphi$ (or $\psi$) and $\vec{\Phi}$ in \reff{Sd}.       

\section{Vacuum solutions: Higgs VEV and corresponding Ricci scalar}

Before we write down the equations of motion for each field in the Einstein action \reff{Sd}, one can readily obtain the Ricci scalar for vacuum solutions. When matter is present, this determines the asymptotic or background spacetime. This will be one of the three maximally symmetric spacetimes: de Sitter space ($R>0$), anti-de Sitter space ($R<0$) or Minkowski spacetime ($R=0$). In the vacua  described by action \reff{Sd}, $\varphi$ is a non-zero constant and the gauge field $A_\mu^i$ is zero. There are two types of vacuum solutions to \reff{Sd}: one with the spontaneous breaking of gauge symmetry where the Higgs field acquires a non-zero VEV and one without. The former can support magnetic monopoles and black holes in all three maximally symmetric spacetimes. However, the unbroken gauge sector supports only black holes. For action \reff{Sd}, moreover, they do not support black holes in a Minkowski background, but only in dS and AdS backgrounds. 

Before we discuss the vacuum solutions to \reff{Sd}, we first describe the case that corresponds to unbroken restricted Weyl symmetry. This case is not accessible to the Einstein action \reff{Sd} but is accessible to the original action \reff{Rw1}. However, we explain why there is not much physical motivation in considering this unbroken sector.

\subsection{Unbroken restricted Weyl symmetry}

In our discussions on the equivalence of the $R^2$ action and the Einstein action, we assumed $\varphi \neq 0$: otherwise the Weyl (conformal) transformation \reff{Conf} is not well-defined and the equivalence does not hold. It is thus important to discuss the case with $\varphi=0$ separately. It turns out that in most cases, the situation actually reduces to the case with $\varphi\neq 0$. This is the case whenever the restricted Weyl symmetry is spontaneously broken either by non-zero $\vec{\Phi}$ or $R$. Let us see how this happens. 

When $\varphi = 0$, the equation of motion for $\varphi$ obtained from action \reff{Sc} demands $R=-\frac{c_2}{\alpha}|\vec{\Phi}|^2$, which still allows non-zero $R$ and $\vec{\Phi}$. In this case, the restricted Weyl invariance is broken, but the conformal transformation \reff{Conf} is invalid. However, there is a simple trick to circumvent this obstruction. We know that these particular values of $R$ and $\vec{\Phi}$ are solutions of the original equations of motion, so one may simply introduce different $c_2$ as $\tilde{c}_2$ because $c_2$ is arbitrary and then $\varphi = \frac{1}{c_1}(-R - \frac{\tilde{c}_2}{\alpha}|\vec{\Phi}|^2)$ becomes non-zero, and we can perform the conformal transformation. In other words, the equivalence to the Einstein action still holds in this case but with different $c_2$.
 
Therefore, the truly exceptional case is when both $R=0$ and $\vec{\Phi} =0$, about which we would like to comment here. Indeed, in the original action \reff{Rw1}, there is a Minkowski background solution with $\vec{\Phi}=0$; this corresponds to a vacuum with $\varphi=0$ in \reff{Sc} and cannot be converted to the case of $\varphi \neq 0$ by changing $c_2$. This vacuum does not break the restricted Weyl symmetry.

The vacuum with unbroken restricted Weyl symmetry may support interesting solutions.
For example, the Schwarzschild blackhole is a solution to the original action \reff{Rw1} with an $R=0$ and $\vec{\Phi}=0$ background. 
However, as already noted, linearizations about Minkowski spacetime (with $\vec{\Phi}=0$) does not yield a propagating graviton but only a propagating scalar (see \cite{Lust3} for details). In other words, Minkowski spacetime does not gravitate in this case \cite{Lust3}.  Thus, we are not going to discuss this case any further for lack of physical motivation.

\subsection{Unbroken gauge sector}     

The $\vec{\Phi}=0$ vacuum with no spontaneous breaking of gauge symmetry yields a cosmological constant of $\Lambda=-c_1/4$ in action \reff{Sd}. The Ricci scalar is then given by 
\begin{align}
R= 4 \,\Lambda=-c_1 \,.
\end{align}
The possible background spacetimes are then de Sitter space if $c_1<0$ and anti-de Sitter space for $c_1>0$. The constant $c_1$ cannot be identically zero and Minkowski space is not a valid background here\footnote{The constant $c_1$ can be made arbitrarily small (but not identically zero) while keeping $\alpha\,c_1$ fixed. In this limiting procedure, one has either dS or AdS spacetime depending on whether $c_1$ is negative or positive respectively. Minkowski spacetime corresponds to $c_1$ being identically zero. To obtain the finite Newton constant by taking $c_1 \to 0$ limit, one has to make $\alpha$ infinity. For finite $\alpha$ the Newton constant is zero, which is another indication that the Minkowski solution with fixed $\alpha$ is not gravitating.
}. This supports black holes in either dS or AdS spacetime only and no magnetic monopoles.

\subsection{Broken gauge sector}

The case with spontaneous breaking of gauge symmetry yields the vacuum solution 
\beq
|\vec{\Phi}|^2=\dfrac{-\alpha\,c_1 \,\xi}{\xi\,c_2-2\alpha\lambda}\quad;\quad R= \dfrac{2\alpha\,c_1 \,\lambda}{\xi\,c_2-2\alpha\lambda}\,.
\eeq{Gauge}
The VEV of $\vec{\Phi}$ breaks the SU(2) gauge symmetry down to $U(1)$, and this breaking pattern is necessary for the existence of the monopole. By recalling $\alpha c_1 <0$ to obtain the conventional sign for the Newton constant, the positivity of $|\vec{\Phi}|^2$ implies that $\xi\,c_2-2\alpha\lambda$ is positive when $\xi>0$ and negative when $\xi<0$. The equation for $R$ then yields anti-de Sitter space when $\xi>0$ and de Sitter space when $\xi<0$. 

A crucial point is that when $\lambda=0$, we obtain an $R=0$ vacuum corresponding to a Minkowski background with $|\vec{\Phi}|^2 = \tfrac{-\alpha\,c_1}{c_2}$ where positivity implies $c_2>0$. In contrast to the Minkowski background solution that exists in the original action \reff{Rw1} with $\vec{\Phi}=0$ (see section 3.1), the $\vec{\Phi}\ne 0$ Minkowski solution is a perfectly viable background and we know linearizations lead to gravitational waves since it is nothing other than the Minkowski space of Einstein gravity. Thus, besides yielding magnetic monopoles, a non-zero VEV for the Higgs field provides $R^2$ gravity with a viable Minkowski background. 

\section{Equations of motion for static spherical symmetry: magnetic monopoles and black holes}

We now look for static spherically symmetric solutions to the final action \reff{Sd}. In this paper, we restrict ourselves to solutions with a fixed non-zero constant for $\varphi$ and we set $\varphi=1$. The equations of motion for $\varphi$ are then satisfied only if the interaction term between $\varphi$ and the Higgs field in \reff{Sd} is set to zero. This requires $6(\xi+2c_2)=1$ as discussed in our previous work \cite{YNAE2}.
 
The ansatz for the metric is
\begin{align}
ds^2 = -B(r) dt^2 + \frac{dr^2}{A(r)} + r^2 d\theta^2 + r^2\sin(\theta)^2 d\phi^2 \ .
\end{align}

Let us make the ansatz for the gauge field
\begin{align}
A^{ia} = q(r) \epsilon^{aik}x^k 
\end{align}
and the Higgs field
\begin{align}
\vec{\Phi} = f(r) \frac{\vec{x}}{r} \ .
\end{align}
Defining 
\begin{align}
1 + r^2 q(r)= a(r)
\end{align}
we have
\begin{align}
D_\mu \vec{\Phi} D^\mu \vec{\Phi} = A\, (f')^2 + 2\frac{a^2f^2}{r^2}
\end{align}
\begin{align}
F_{\mu\nu}F^{\mu\nu} = \frac{4 A \,(a')^2}{r^2}  + \frac{2(a^2-1)^2}{r^4}
\end{align}
where a prime denotes derivative with respect to $r$.
The final action \reff{Sd} becomes 
\begin{align}
S &= \int d^4x \sqrt{-g}(\tilde{\Lambda} + \tilde{\alpha} R - \tilde{\xi} R \vec{\Phi}^{\,2}  - D_\mu \vec{\Phi} D^\mu \vec{\Phi} -\tilde{m}^2 \vec{\Phi}^2 - \tilde{\lambda} (\vec{\Phi}^2)^2 - \frac{1}{4g^2}F_{\mu\nu}^2 \cr
 &= 4\pi \int dr dt  \small{\sqrt{B/A}}\, r^2 \bigg[\tilde{\Lambda} - \tilde{m}^2 f^2 -A (f')^2-\dfrac{2 a^2 f^2}{r^2} -\tilde{\lambda} f^4 
-\dfrac{(a^2-1)^2}{2g^2r^4} - \dfrac{A (a')^2}{g^2r^2} \cr
&\qquad\qquad+(\tilde{\alpha}-\tilde{\xi}f^2)\Big(\dfrac{2}{r^2}-\dfrac{2 A}{r^2}-\dfrac{2 A'}{r}-\dfrac{2 A B'}{r B}-\dfrac{A' B'}{2 B}+\dfrac{A (B')^2}{2 B^2}-\dfrac{A B''}{B}\Big)\bigg]
\end{align}
where for convenience we introduced the new parameters 
\beq
\tilde{\Lambda}= -\alpha c_1^2\quad;\quad \tilde{\alpha}=-2\alpha c_1 \quad;\quad\tilde{\xi}=\xi + 2 c_2\quad;\quad\tilde{m}^2=2 c_1c_2\quad;\quad \tilde{\lambda}=\lambda + c_2^2/\alpha \,.
\eeq{NP}
Note that  $6(\xi+2c_2)=1$ fixes $\tilde{\xi}$ to be $1/6$; the other parameters are arbitrary.
Varying $A$,$B$,$f$,$a$, we obtain the following equations of motion:
\begin{align}
0&=-4 g^2 r^3 A B' (-\tilde{\alpha}+\tilde{\xi} f^2+r \tilde{\xi} f f')+B \Big(1-4 g^2 r^2 \tilde{\alpha}-2 g^2 r^4 \tilde{\Lambda}\cr &\qquad+a^4+2 g^2 \tilde{m}^2 r^4 f^2+4 g^2 r^2 \tilde{\xi} f^2+2 g^2 r^4 \tilde{\lambda} f^4+a^2 (-2+4 g^2 r^2 f^2)\cr&\qquad\qquad\qquad-2 r^2 A \big(2 g^2 \tilde{\xi} f^2+a'^2+8 g^2 r \tilde{\xi} f f'+g^2 (-2 \tilde{\alpha}+r^2 f'^2)\big)\Big).
\label{EOM_A}
\end{align}
\begin{align}
0&=-1+4 g^2 r^2 \tilde{\alpha}+2 g^2 r^4 \tilde{\Lambda}-a^4-2 g^2 \tilde{m}^2 r^4 f^2-4 g^2 r^2 \tilde{\xi} f^2-2 g^2 r^4 \tilde{\lambda} f^4\cr&\qquad+a^2 (2-4 g^2 r^2 f^2)-4 g^2 r^3 \tilde{\alpha} A'+4 g^2 r^3 \tilde{\xi} f^2 A'+4 g^2 r^4 \tilde{\xi} f A' f'\cr&\quad\qquad+2 r^2 A \big(2 g^2 \tilde{\xi} f^2-a'^2-g^2 (2 \tilde{\alpha}+r^2 (1-4 \tilde{\xi}) f'^2)+4 g^2 r \tilde{\xi} f (2 f'+r f'')\big).
\label{EOM_B}
\end{align}
\begin{align}
0&=-r^2 \tilde{\xi} A f B'^2+r B \Big (r \tilde{\xi} f A' B'+A \big(r B' f'+2 \tilde{\xi} f (2 B'+r B'')\big)\Big)+B^2 \Big(-4 r^2 \tilde{\lambda} f^3\cr&\qquad\qquad-2 f (\tilde{m}^2 r^2+2 \tilde{\xi}+2 a^2-2 \tilde{\xi} A-2 r \tilde{\xi} A')+r (4 A f'+r A' f'+2 r A f'')\Big).
\label{EOM_f}
\end{align}
\begin{align}
0&=-2 a^3 B+2 a B (1-2 g^2 r^2 f^2)+r^2 \big(A a' B'+B \big(a' A'+2 A a''\big)\big).
\label{EOM_a}
\end{align}

\subsection{Analytical black hole solutions}

The system of equations \reff{EOM_A}-\reff{EOM_a} admits an analytical solution which is a Reissner-Nordstr\"{o}m (RN) black hole with $a=0$, $f=f_0=\word{constant}$and $A=B=1+ k\, r^2-\mu/r+Q/r^2$ where $k$, $\mu$ and $Q$ are constants. $Q$ will later be expressed in terms of the actual charge $q$ and $k=0$, $k>0$ and $k<0$ correspond to Minkowski, AdS and dS backgrounds respectively. The equations \reff{EOM_A} and \reff{EOM_f} yield the following relations below (whereas equations \reff{EOM_B} and \reff{EOM_a} do not yield any new relations):
\begin{align}
Q\,(\tilde{\alpha}-\tilde{\xi}f_0^2) &= \frac{1}{4 g^2} \label{Q1}\\
\tilde{\Lambda} - 6\,k (\tilde{\alpha} - \tilde{\xi} f_0^2) &= \tilde{m}^2 f_0^2 + \tilde{\lambda} f_0^4 \label{Q2}\\
12 \,k \tilde{\xi} \,f_0 -\tilde{m}^2\,f_0  - 2\tilde{\lambda} f_0^3 &= 0\,. \label{Q3}
\end{align}
Note that $\mu$ is arbitrary and the constants $Q$, $k$ and $f_0$ can be determined in terms of the parameters of the theory, which we do now. 

In the RN metric, $Q$ is identified as $G\,q^2$ where $G$ is Newton's constant and $q$ is the conserved charge. Note also that when $f=f_0$, $\tilde{\alpha}-\tilde{\xi}f_0^2$ is the coefficient of the Ricci scalar in action \reff{Sd}. Therefore we identify $\tilde{\alpha}-\tilde{\xi}f_0^2$ with $1/(16\, \pi\,G)$. Substituting this into \reff{Q1} we obtain a simple formula for the charge $q$: 
\beq
q^2=\dfrac{4 \,\pi}{g^2} \,.
\eeq{charge}

Solving equations \reff{Q2} and \reff{Q3}, we obtain the values of $f_0$ and $k$:
\beq
f_0^2=\dfrac{2 \tilde{\Lambda} \tilde{\xi}-\tilde{m}^2 \tilde{\alpha}}{\tilde{m}^2 \tilde{\xi} + 2 \tilde{\alpha}\tilde{\lambda}} \qquad ;\qquad
k=\dfrac{1}{12}\,\,\dfrac{4 \tilde{\Lambda} \tilde{\lambda}+\tilde{m}^4}{\tilde{m}^2 \tilde{\xi} + 2 \tilde{\alpha}\tilde{\lambda}}\,.
\eeq{fok}
The above equations are in agreement with the vacuum results \reff{Gauge} when the new parameters \reff{NP} are expressed in terms of the original parameters (recall also that $R=-12k$).

Our analytical solution is a generalization to curved backgrounds and non-minimal coupling of the EYMH black hole analytical solution found for asymptotically flat spacetime and minimal coupling \cite{Cho}.

\section{Magnetic Monopoles: numerical solutions}
We numerically seek gravitating magnetic monopole solutions to the equations of motion \reff{EOM_A}-\reff{EOM_a} in flat, dS and AdS backgrounds with non-minimal coupling constant $\tilde{\xi}=1/6$. By definition, magnetic monopoles are non-singular at the origin $r=0$ and have a field configuration of a point-like magnetic charge at large distances. This requires in total five boundary conditions at $r=0$ and ``infinity":
\beq
a(0)=1 \quad;\quad A(0)=1 \quad;\quad f(0)=0 \quad ;\quad a(\infty)=0 \quad;\quad f(\infty)=1\,.
\eeq{boundary}
There are five and not seven boundary conditions because the metric function $B(r)$ can be eliminated from the equations of motion. This is accomplished by extracting $W=B'/B$ from equation \reff{EOM_A} and  substituting it in equations \reff{EOM_f} and \reff{EOM_a}. One must also substitute $B''/B= W' +W^2$ into \reff{EOM_f}. This reduces the number of equations from four to three and the number of initial conditions from seven to five since we do not need to specify the initial conditions for $B(r)$ and $B'(r)$. Note that $W$ determines $B$ up to an overall rescaling of $B$, which can always be absorbed by the definition of the time $t$.

We plot the functions $f(r)$ and $a(r)$ corresponding to magnetic monopoles for three different backgrounds: Minkowski, AdS and dS. The common parameters used in all three plots are: $g=1$, $\tilde{\alpha}=10$, $\tilde{\lambda}=1 \times 10^{-3}$ and $\tilde{\xi}=1/6$. The two parameters $\tilde{m}^2$ and $\tilde{\Lambda}$ differ in all three backgrounds as they depend on the value of the background scalar curvature $R=-12k=4\Lambda$ where $\Lambda$ (distinct from $\tilde{\Lambda}$) is the cosmological constant. We specify $\Lambda$ (or $k=-\Lambda/3$) and obtain the two parameters $\tilde{m}^2$ and $\tilde{\Lambda}$ from equations \reff{Q2} and \reff{Q3} with $f_0=1$\footnote{Note that equations \reff{Q1}, \reff{Q2} and \reff{Q3} are not only valid for  the RN black hole but also for the magnetic monopole as they are equations corresponding to the asymptotic limit.}. For Minkowski space, we plot the metric function $A(r)$ but for AdS and dS space we plot $(A(r)-1)/r^2$ instead since $A(r)$ has the asymptotic form $1+ k r^2$ and  $(A(r)-1)/r^2$ plateaus to the value of $k$.

\pagebreak
\begin{figure}[h]
 \caption{\label{Flat}Monopole in flat background with non-minimal coupling. This case corresponds to $k=0$. }
 \includegraphics[scale=1]{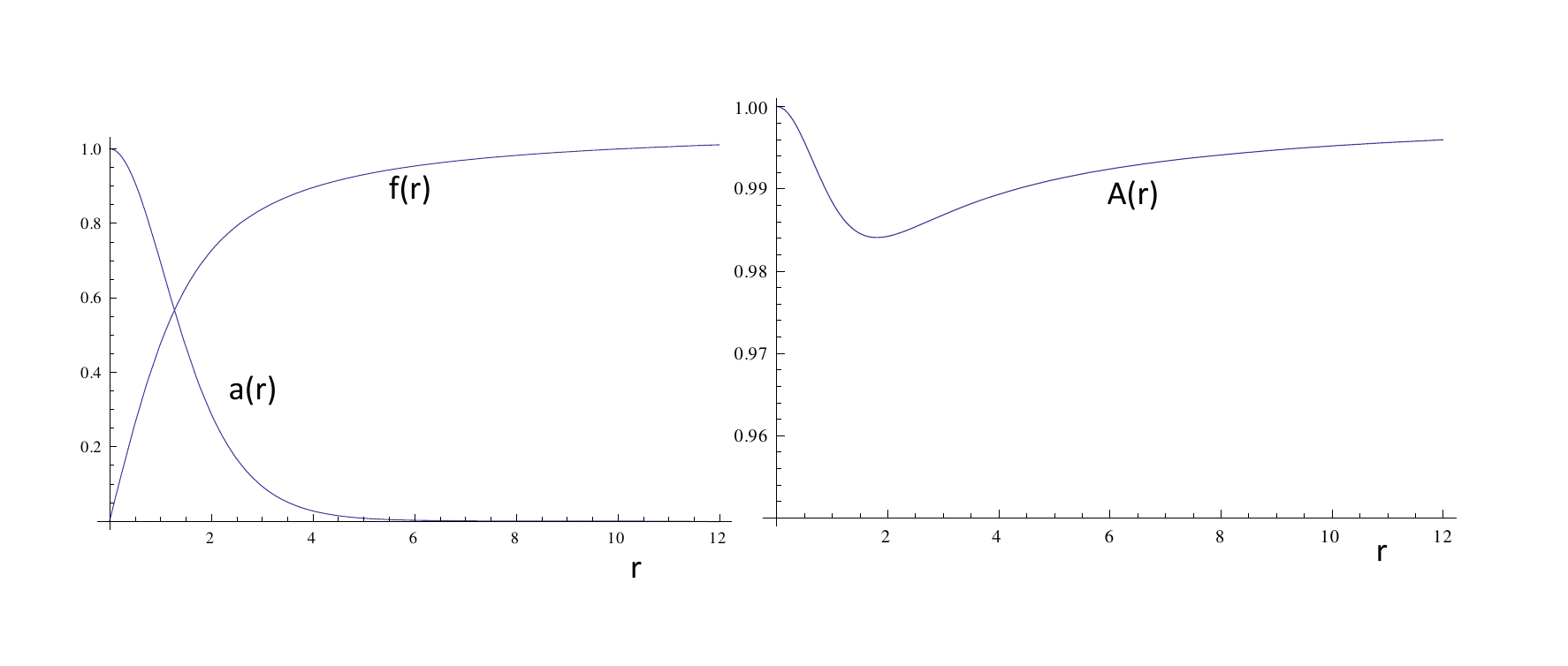}
 \end{figure}	   
\pagebreak              
\begin{figure}[h]
 \caption{\label{AdS}Monopole in AdS background with non-minimal coupling. The cosmological constant is chosen to be $\Lambda=-1$ (hence $k=1/3$).}
 \includegraphics[scale=1]{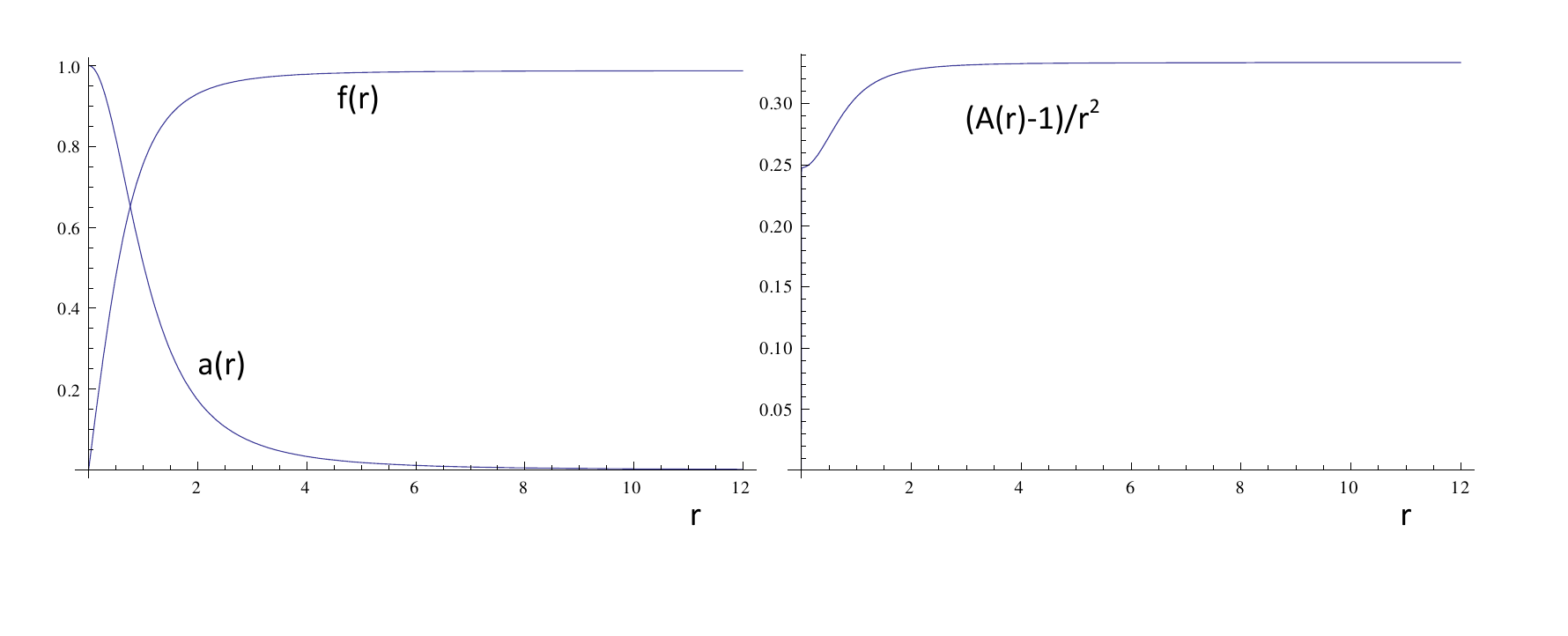}
 \end{figure}	  

\pagebreak
\begin{figure}[h]
 \caption{\label{dS}Monopole in dS background with non-minimal coupling. The cosmological constant is $\Lambda=+0.005$ (hence $k=-0.001667$).}
 \includegraphics[scale=1]{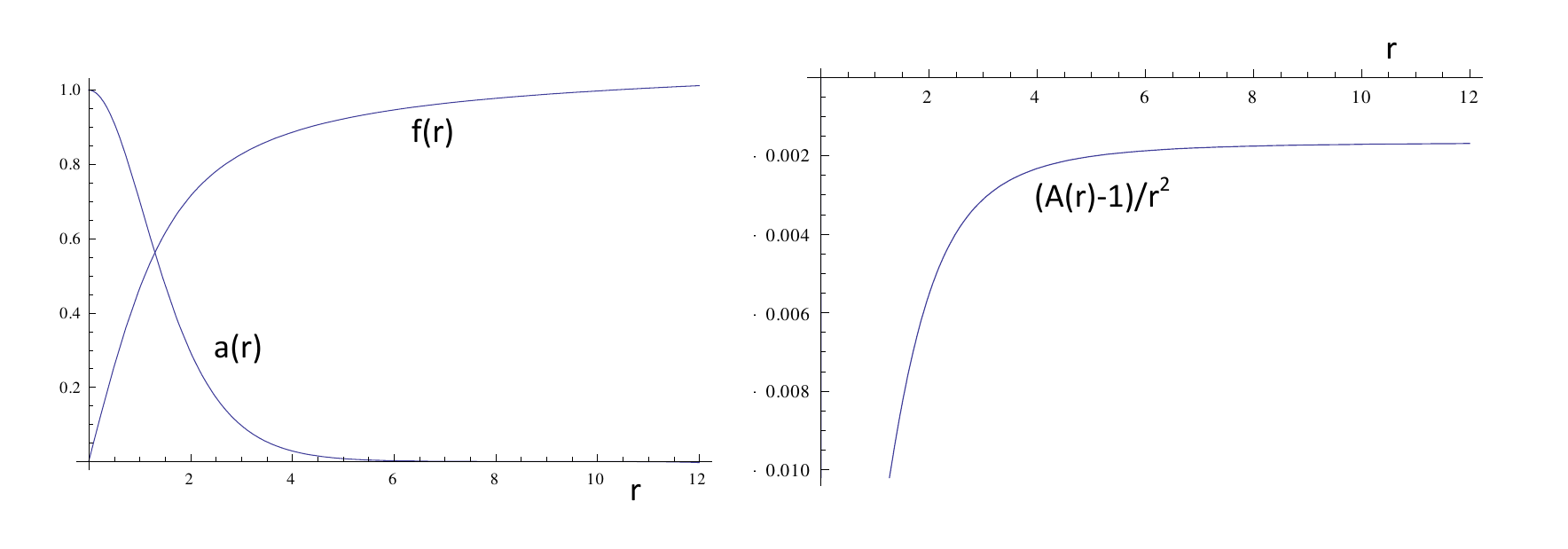}
 \end{figure}	  

\section{Conclusion}

In this paper we obtained numerically gravitating magnetic monopole solutions from the spontaneous symmetry breaking of an original restricted Weyl invariant action containing pure $R^2$ gravity and a non-minimally coupled massless Higgs field together with Yang-Mills fields. Our solutions are part of the EYMH monopoles but with a non-zero coupling constant $\tilde{\xi}=1/6$ and take place in three different backgrounds: Minkowski, dS and AdS. One of the important spin-offs of this work is that although the unbroken sector of the original theory does not have a viable Minkowski background because linearizations about it do not yield gravitons, in the broken gauge sector, the Minkowski background is gravitating and perfectly fine. Note that the broken gauge sector exists only if the coupling constant $\xi$ in the original action \reff{Rw1} is non-zero. Therefore the non-minimal term $\xi R |\vec{\Phi}|^2$ in the original action has two consequences: it allows for magnetic monopole solutions but just as important, it gives the $R^2$ theory a viable Minkowski background.

In this work we set $\varphi$ to unity and satisfied the equations of motion of $\varphi$ by eliminating the interaction term between $\varphi$ and the Higgs field by choosing $\tilde{\xi}=1/6$. This was done in order to focus on the EYMH monopoles which have no extra scalar $\varphi$ and hence no such interaction term. Nonetheless, the interaction term is a distinctive part of the theory and could be the origin of a new interaction/force and experimentally testable.
 It would be therefore interesting and important to investigate how it affects the results in future work.

\pagebreak

\section*{Acknowledgments}
A.E. acknowledges support from a discovery grant of the National
Science and Engineering Research Council of Canada (NSERC).  
Y.N. is supported in part by JSPS KAKENHI Grant Number 17K14301.

\end{document}